\documentclass[a4paper,conference]{IEEEtran}
\ifCLASSINFOpdf
\else
\fi
\usepackage{amsmath,amssymb,amsfonts}
\usepackage{algorithmic}
\usepackage{graphicx}
\usepackage{textcomp}
\usepackage{multirow}
\usepackage{multicol}
\usepackage{hyperref}
\usepackage{hyperref}

\usepackage{graphicx}

\usepackage{xcolor}

\begin{document}
%
\title{CVAD: A Generic Medical Anomaly Detector Based on Cascade VAE}



\author{\IEEEauthorblockN{Xiaoyuan Guo\IEEEauthorrefmark{1},
Judy Wawira Gichoya\IEEEauthorrefmark{2,3},
Saptarshi Purkayastha\IEEEauthorrefmark{4}, and
Imon Banerjee\IEEEauthorrefmark{5}}
\IEEEauthorblockA{\IEEEauthorrefmark{1}Department of Computer Science, Emory University, Georgia, USA}
\IEEEauthorblockA{\IEEEauthorrefmark{2}Department of Radiology and Imaging Sciences, Emory University, Georgia, USA}
\IEEEauthorblockA{\IEEEauthorrefmark{3}Department of Biomedical Informatics, Emory University, Georgia, USA}
\IEEEauthorblockA{\IEEEauthorrefmark{4}School of Informatics and Computing, Indiana University-Purdue University Indianapolis, IN, USA}
\IEEEauthorblockA{\IEEEauthorrefmark{5}School of Computing, Informatics, and Decision Systems Engineering, Arizona State University, AZ, USA\\ Email: \{xiaoyuan.guo,judywawira\}@emory.edu, saptpurk@iupui.edu, banerjee.imon@mayo.edu}}


\maketitle

\begin{abstract}
 Detecting out-of-distribution (OOD) samples in medical imaging plays an important role for downstream medical diagnosis. However, existing OOD detectors are demonstrated on natural images composed of classes with clear inter-class variations and have difficulty generalizing to medical images. The key issue is the granularity of OOD data in the medical domain, where intra-class OOD samples are predominant. We focus on the generalizability of OOD detection for medical images and propose a self-supervised \textbf{C}ascade \textbf{V}ariational autoencoder-based \textbf{A}nomaly \textbf{D}etector (CVAD). We use a cascaded variational autoencoder architecture, which combines latent representation at multiple scales, before being fed to a discriminator to distinguish the OOD data from the in-distribution (ID) data. Finally, both the reconstruction error and the OOD probability predicted by the binary discriminator are used to determine the anomalies. We compare the performance with the state-of-the-art deep learning models to demonstrate our model's efficacy on various open-access medical imaging datasets for both intra- and inter-class OOD. Further extensive results on datasets including common natural datasets show our model's effectiveness and generalizability.
\end{abstract}

%
\IEEEpeerreviewmaketitle

\section{Introduction}
\begin{figure}[tp]
\centering
\includegraphics[width=0.45\textwidth]{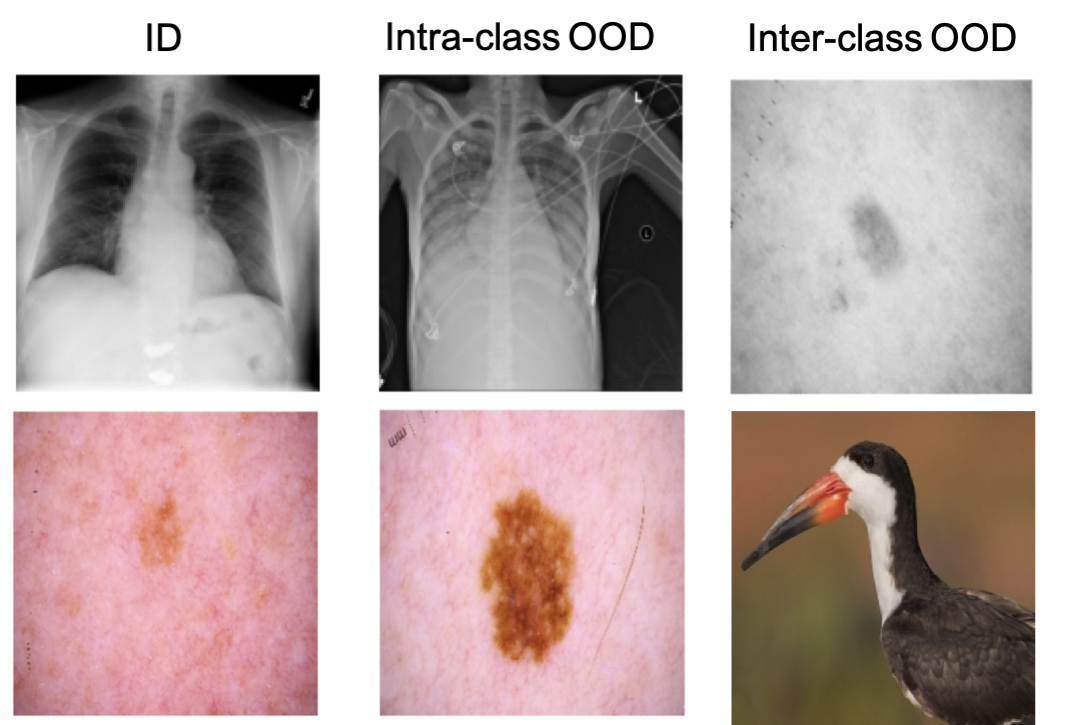}
\caption{ID, Intra- and Inter-class OOD examples for medical images. Compared to natural images, medical OOD samples exhibit more subtle intra-class variations (e.g., normal vs pneumonia in the 1st row and benign vs malignant in the 2nd row).} 
\label{fig_intra_inter}
\vspace{-5mm}
\end{figure}

Despite recent advances in deep learning that have contributed to solving various complex real-world problems~\cite{daxberger2019bayesian,duan2021bridging}, the safety and reliability of AI technologies remain a big concern in medical applications. Deep learning models for medical tasks are often trained with data from known distributions, and fail to identify out-of-distribution (OOD) inputs and possibly assign high probabilities to the anomalies during inference because of the insensitivity to distribution shifting. Medical anomalies, \textit{a.k.a., OOD data, outliers}, can arise due to various reasons such as noise during data acquisition, changes in disease prevalence and incidence (\textit{e.g.}, the evolution of rare cancer types), or inappropriate inputs (\textit{e.g.}, different modalities unseen during training)~\cite{fernando2020deep}. To ensure the reliability of deep models' predictions, it is necessary to identify unknown types of data that are different from the training data distribution. A good anomaly detector should be able to learn good representations of the in-distribution (ID) during training and thus identify the outliers from test datasets. However, the core challenges for medical anomaly detection are -- (1) the OOD data is usually unavailable at the time of model training; (2) in theory, there are infinite numbers of variations of OOD data; and (3) different types of OOD data can be identified with varying difficulties. In general, the OOD classifications~\cite{cao2020benchmark} can be refined based on the variation difference by summarizing them as \textbf{inter-class} OOD data and \textbf{intra-class} OOD data. Inter-class OOD data is in a category different from the ID data\footnote{By default, we mean a category can contain several classes. For example, a bird category can include owls, woodpeckers, flamingos, etc. }, e.g. a skin cancer image v.s. a lung X-ray image; intra-class OOD data belongs to the same category as the ID data but different classes, e.g. a normal skin image v.s. a skin image with cancer. Therefore, inter-class OOD data often has larger variations from the ID data, whereas the intra-class OOD data is close to ID data, as observed in Figure~\ref{fig_intra_inter}. Thus, identifying intra-class OOD data is more difficult than the inter-class OOD data given subtle differences with ID data.

To cope with the OOD unavailability and uncertainty challenges, we adopt an unsupervised way to design our anomaly detector. For intra-class OOD data, we expect the model can be sensitive to minor variations and thus screen the dissimilar inputs. To acquire such high identification of hard OOD cases, we propose a \textbf{C}ascade \textbf{V}ariational autoencoder based \textbf{A}nomaly \textbf{D}etector (CVAD) to learn both coarser and finer features inspired by~\cite{larsen2016autoencoding,bao2017cvae}.  With the cascade VAE architecture to model the in-distribution representations, CAVD gains superior reconstructions and learns good-quality features to threshold out the OOD data. To enhance the detection ability of inter-class OOD data, we further train a binary discriminator with the reconstructed data as the fake OOD category. In this paper, our contributions are three-fold: 
\begin{itemize}
    \item We propose a generic medical OOD detector -- CVAD. By utilizing a cascade VAE to learn latent variables of in-distribution data, CVAD owns good reconstruction ability of in-distribution inputs and obtains discriminative ability for OOD data based on the reconstruction error.
    \item We adopt a binary discriminator to further separate the in-distribution data from the OOD data by taking the reconstructed image as fake OOD samples. Thus, our model has better discriminative capability for the inter-class as well as intra-class OOD cases.
    \item We conduct extensive experiments on multiple public medical image datasets to demonstrate the generalization ability of our proposed model. We evaluate comprehensively against state-of-the-art anomaly detectors in detecting both intra-class and inter-class OOD data, showing improved performance. The implementation technical report including original code and usage instructions has been publicly available in~\cite{guo2021cvad}.
\end{itemize}

\section{Related Work}

Although there have been extensive research on outlier detection~\cite{daxberger2019bayesian,abati2019latent}, effective medical image OOD detectors are still lacking due to complicated data types (e.g., various modalities and protocols, difference in acquisition devices) and user-defined application situations (e.g., disease types). Without OOD data available during training, unsupervised anomaly detection becomes the mainstream research direction, which CVAD also belongs to. Recent unsupervised anomaly detection approaches can be roughly classified as two main categories - generative and objective.

\subsection{Generative methods}
Deep generative models appear to be promising in detecting OOD data since they can learn latent features of training data and generate synthetic data with similar features to known classes~\cite{li2018anomaly}. Thus, the compressed latent features can be used to distinguish OOD data from ID data. Two major families of deep generative models are Variational Autoencoders (VAEs)~\cite{kingma2013auto} and Generative Adversarial Networks (GANs)~\cite{goodfellow2014generative}. 

\textbf{VAEs:} Traditional AutoEncoders~\cite{mcclelland1986parallel} (AEs) can reconstruct input images well and be used to detect anomalies~\cite{baur2021autoencoders}, but risk learning the identity of deep image features. Comparatively, VAEs generate contents by regularizing the latent feature distribution representations. With this trait, VAE~\cite{kingma2013auto} and its modifications have been used widely in generating realistic synthetic images~\cite{zimmerer2019high,huang2018introvae,sonderby2016ladder}. Although VAEs are theoretically elegant and easy to train with nice manifold representations, they usually produce blurry images that lack detailed information~\cite{bao2017cvae,huang2018introvae}. To improve the image reconstruction quality, pchVAE~\cite{zimmerer2019high} adds a conditional hierarchical VAE branch to learn lower-level image components. The improved reconstructions of VAEs are adopted for detecting OOD samples based on the reconstruction quality~\cite{an2015variational}. Other approaches seek to enhance the reliable uncertainty estimation of VAE for better performance~\cite{daxberger2019bayesian,xiao2020likelihood,ran2020detecting,pol2019anomaly,somepalli2020unsupervised}. Reference~\cite{ran2020detecting} applies an improved noise contrastive prior (INCP) to acquiring reliable uncertainty estimate for standard VAEs; whereas Bayesian VAE~\cite{daxberger2019bayesian} detects OOD by estimating a full posterior distribution over the decoder parameters using stochastic gradient Markov chain Monte Carlo. Nonetheless, most of the VAE-based OOD detections are only evaluated on natural image datasets (MNIST~\cite{lecun-mnisthandwrittendigit-2010}, FashionMNIST~\cite{xiao2017fashion}, CIFAR10~\cite{krizhevsky2010cifar}, SVHN~\cite{netzer2011reading}, \textit{etc.}), which are with small image size (e.g., $32\times32$) and clear intra- and inter-class variations. 

\begin{figure*}[!htp]
\centering
\includegraphics[width=0.9\textwidth]{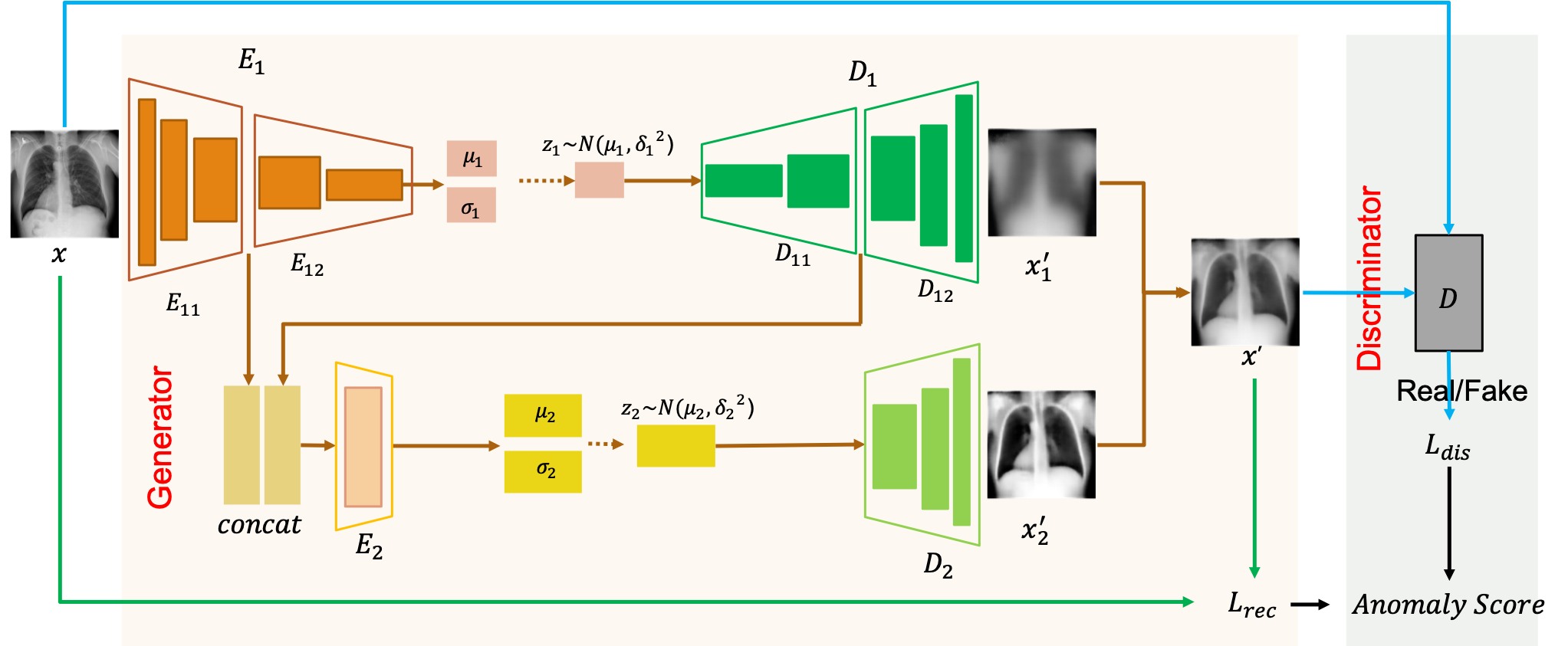}
\caption{Proposed CVAD architecture - a cascade VAE as the generator and a separate binary classifier (D) as the discriminator. The main VAE pipeline is composed by the encoder $E_{1}$ shown as the orange part and the decoder $D_{1}$ in the dark green part; the branch VAE has the pink part as the encoder $E_{2}$ and the light green for its decoder $D_{2}$. Given an input image $x$, the main VAE learns to reconstruct $x^{'}_{1}$ via latent representations $\mu_{1}$ and $\sigma_{1}$; the branch VAE takes the outputs of the results of the main VAE encoder intermediate part $E_{11}$ and the intermediate decoder $D_{11}$ as inputs and feeds the concatenated features to $E_{2}$ to formulate the branch latent variables $\mu_{2}$ and $\sigma_{2}$, which gives a low-level reconstruction $x^{'}_{2}$ via the corresponding decoder $D_{2}$. By adding the two reconstructions - $x^{'}_{1}$ and $x^{'}_{2}$together with a sigmoid function, a final reconstruction $x$ is generated and later treated as fake OOD data as compared to the original input $x$. The binary discriminator $D$ will learn to distinguish them.} 
\label{fig_cvad}
\vspace{-5mm}
\end{figure*} 

\textbf{GANs:} Compared with VAEs, GANs usually generate much sharper images but face challenges in training stability and sampling diversity, especially when synthesizing high-resolution images~\cite{huang2018introvae}. Still, GANs remain popular in outlier detection, such as, AnoGAN~\cite{schlegl2017unsupervised}, f-AnoGAN~\cite{schlegl2019f}, ADGAN~\cite{deecke2018image}, GANomaly~\cite{akcay2018ganomaly} to detect OOD samples using GAN architectures. Besides standard architectures, there are hybrid models that detect anomalies by combining a VE/VAE with a GAN~\cite{larsen2016autoencoding,bao2017cvae,perera2019ocgan}. In order to acquire competitive OOD discriminative ability, OCGAN~\cite{perera2019ocgan} integrates four components: a denoising auto-encoder, two discriminators and a classifier with complicate training process. Generally, such hybrid networks are not competitive for image datasets with clear class variations, as reported in~\cite{larsen2016autoencoding}. Their experiments are often done with small-sized images and may fail when experimenting on large-sized medical images.

\subsection{Objective methods}
Objective anomaly detectors learn identifying OOD data via specific optimization functions and auxiliary transformations. Such OOD detection approaches include classifier-based and transformation-based methods~\cite{ruff2018deep,tack2020csi,liang2017enhancing}. 

\textbf{Classifier-based method:} ODIN~\cite{liang2017enhancing} uses temperature scaling and adds small perturbations to input data for separating the softmax score distributions between ID and OOD images. Similar separation via a multi-class classifier is also followed by~\cite{yu2019unsupervised}. However, the prerequisite of balanced multiple classes is not always applicable in medical applications. Comparatively, the one-vs-rest setup~\cite{liznerski2020explainable:explainable} is much more common and useful in medical OOD detection, which treats one-class as in-distribution data and evaluates performance on the left OOD data. Following the setting, the anomaly detection reduces to a one-class classification (OCC) problem~\cite{khan2014one}. Representative one-class classifiers are DeepSVDD~\cite{ruff2018deep}, OCSVM~\cite{scholkopf2001estimating}. 

\textbf{Transformation-based method:} Most of the anomaly detectors are unsupervised given the assumption the anomalies are unavailable during training. Hence, good detection performance largely depends on the learning of high-quality in-distribution features. Self-augmentation with transformations on training data not only enriches the training diversity but also introduces discriminative knowledge. For example, ~\cite{tack2020csi} proposes contrasting shifted instances for anomaly detection. Nevertheless, the augmentations are with limited transformations and consume more time to train as more generated data are fed as fake OOD data. Our model CVAD has no additional augmentations but still captures high-quality representations of in-distribution data.   
Besides, there are many other approaches contributing to OOD detection, such as GradCon~\cite{kwon2020backpropagated}, generalized ODIN~\cite{hsu2020generalized} and FSSD~\cite{huang2020feature}. Please refer to the papers for more details.

\section{Methods}

Anomaly detection includes both intra- and inter-class OOD identification, of which medical intra-class OOD data is much more challenging because of the minute dissimilarity compared to ID data. With no prior knowledge available and no sophisticated pre-processing, we utilize a variational autoencoder to learn the ``normality" of in-distribution inputs via image reconstruction and enhance the discriminative ability for both two OOD classes via a binary discriminator. Both the reconstruction and discrimination contribute to accurate intra- and inter-class OOD detection. 

\subsection{CVAD architecture}\label{arch}
Figure~\ref{fig_cvad} shows the design of CVAD. Inspired by the GAN's architecture, we adopt a cascade VAE architecture as the ``generator" for modeling ID representations and a separate classifier as the ``discriminator" to strengthen OOD discrimination. 
 
A standard VAE module consists of two neural networks: an encoder and a decoder~\cite{kingma2013auto}, with the encoder $q_{\phi}(z|x)$ (parameterized by $\phi$) mapping the visible variables $x$ to the latent variables $z$ and the decoder $p_{\theta}(x|z)$ (parameterized by $\theta$) sampling the visible variables $x$ given the latent variables $z$~\cite{huang2018introvae}. 
Given a dataset $D={\{x_{i}\}^{N}_{i=1}}$ with $N$ input vectors drawn from some underlying data distribution $p^{*}(x)$, $\phi$ and $\theta$ are then learned by maximizing the variational lower bound (ELBO) $L(\phi,\theta)$, which is a lower bound to the marginal log-likelihood $\log p (x|\theta)$~\cite{daxberger2019bayesian}. However, a vanilla VAE exhibits limited potential in distinguishing unseen distributions due to the blurry reconstructions for large-size images. Thus, we learn from pchVAE~\cite{zimmerer2019high} and tailored it as ``generator" to acquire high-quality reconstruction and better latent representations.

\textbf{Generator:} Different from the standard VAE, our ``generator" has two encoders $E_{1}, E_{2}$ and two decoders $D_{1}$, $D_{2}$. To learn the high-level features, a deep and standard VAE architecture constructed by $E_{1}$ and $D_{1}$ formulates the deep latent variables $z_{1}$ by sampling parameters $\mu_{1}$ and $\sigma_{1}$ of size $K$. Meanwhile, the low-level features are learnt by the branch VAE. Instead of using the original input, branch VAE utilizes the concatenation of two intermediate features from $E_{11}$ and $D_{11}$. Given original input variables $x$, the input of branch VAE can be represented as $f(x)$. The encoder of branch VAE $E_{2}$ is simpler than $E_{1}$ whereas the decoder $D_{2}$ owns the same architecture as $D_{12}$. This branch VAE formulates latent Gaussian distributions with parameters $\mu_{2}, \sigma_{2}$ of size $4K$. After sampling, two sets of latent variables, i.e., $z_{1}, z_{2}$ are acquired and decoded to image contexts $x_{1}^{'}$ and finer details $x_{2}^{'}$ respectively. $x$ is the combination of $x_{1}^{'}$ and $x_{2}^{'}$. 

\textbf{Discriminator:} Since the ``generator" itself has no awareness of distinguishing outliers, we add a binary discriminator $D$ to distinguish the reconstructed image $x^{'}$ from the original input image $x$. As $x^{'}$ shares very similar features with $x$ after the first-stage training of the image generator, the discriminator is much more sensitive to minor differences from the in-distribution data, enhancing the accuracy of identifying both intra-class OOD data and inter-class OOD data.

\subsection{Network training}
Instead of training CVAD in an adversarial way, we train the generator and the discriminator in two stages. The reason is that training with adversarial losses often leads to much sharper reconstructions but ignores the low-level information of ID data, incurring high reconstruction errors and potential dangerous decisions for medical applications. Therefore, CAVD is designed to first train the image generator and then the binary discriminator to detect OOD data. This non-adversarial training enables CVAD to inherit the merit of VAEs~\cite{kingma2013auto} and avoid the instability of GANs~\cite{goodfellow2014generative}.

To optimize the generator, we minimize two objectives for the primary VAE part in Eqn.~\ref{vae1} and the branch VAE part in Eqn.~\ref{vae2}, KL refers to Kullback-Leibler divergence.  
\begin{equation}\label{vae1}
\begin{split}
L(x;\phi_{1},\theta_{1}) =
-E_{z_1\sim q_{\phi_1}{(z_{1}|x)}}[{log}\: p_{\theta_1}(x|z_{1})] + \\
D_{KL}(q_{\phi_{1}}(z_{1}|x)||p_{\theta_1}(z_{1}))
\end{split}
\end{equation}
\begin{equation}\label{vae2}
\begin{split}
L(x;\phi_{2},\theta_{2})=-E_{z_2\sim q_{\phi_2}{(z_{2}|f(x))}}[{log}\: p_{\theta_2}(x|z_{2})] + \\
D_{KL}(q_{\phi_{2}}(z_{2}|f(x))||p_{\theta_2}(z_{2}))
\end{split}
\end{equation}
Therefore, the ``generator" loss can be formulated as Eqn.~\ref{cvae}. $\alpha_1$ and $\alpha_2$ to balance the weights of the two individual terms.
\begin{equation}\label{cvae}
L_{\rm{G}} = \alpha_1 L(x;\phi_{1},\theta_{1}) + \alpha_2 L(x;\phi_{2},\theta_{2})
\end{equation}
The binary discriminator is trained to distinguish true/fake images using binary cross entropy. 

\textbf{Anomaly score:}
An anomaly score $S$ is defined in Eqn.~\ref{score} based on errors during inference and includes two parts: the reconstruction error $L_G$ output by the ``generator" and the probability of being the anomaly class $S_D$ output by the discriminator. Instead of simply adding the two parts together, we first scale the ``generator" reconstruction errors into [0,1] for the whole dataset and get the average score value to avoid assigning imbalanced weights between the two parts: 
\begin{equation}\label{score}
S = 0.5*({\frac{{L}_{G}-{L_{G}}_{{min}}}{{L_{G}}_{{max}}-{L_{G}}_{{min}}}} + S_{D})
\end{equation}

\subsection{Network Details}
As illustrated in Figure~\ref{fig_cvad}, our generator has a standard VAE part which consists of $E_{11}$, $E_{12}$, $D_{11}$ and $D_{12}$ and a branch VAE composed by a shallow encoder $E_{2}$ and a decoder $D_{2}$. The primary VAE is a symmetric network with five $4\times4$ convolutions with stride 2 and padding 1 followed by five transposed convolutions. Respectively, $E_{11}$ stands for the first three convolution layers; $E_{12}$ refers the last two convolution layers; $D_{11}$ is for the first three transposed convolution layers and $D_{12}$ means the last two transposed convolution layers. The input of the branch VAE is the intermediate features of $E_{11}$ and the middle decoded features of $D_{11}$. $E_{2}$ here is a convolution layer which has a same $4\times4$ kernel with stride 2 and padding 1. $D_{2}$ shares the same decoder architecture as the standard VAE, namely, $D_{2}=D_{11}+D_{12}$. All convolutions and transposed-convolutions are followed by batch normalization and leaky ReLU (with slope 0.2) operations. We used a base channel size of 16 and increased number of channels by a factor of 2 with every encoder layer and decreased the number of channels to half for each decoder layer. The latent dimension $K$ of $z_{1}$ is set as 512 and $z_{2}$ is with $4K$, i.e., 2048 dimensions.  
The binary discriminator is composed of five convolution layers with the same settings as above and a final fully connected layer to make a binary prediction. After a sigmoid function, the final ID/OOD class probability is obtained. 

\begin{table}[tp]
\caption{The selection details of ID and OOD data\label{details}}
\centering
\resizebox{0.38\textwidth}{!}{%
\begin{tabular}{|l|l|}
\hline
\textbf{Dataset} & \textbf{Details} \\ \hline
\multirow{5}{*}{RSNA} & \textit{In-class:} normal (8,851) \\ \cline{2-2} 
 & \begin{tabular}[c]{@{}l@{}}\textit{Intra-class:} pneumonia (9,555),\\  abnormal (11,821)\end{tabular} \\ \cline{2-2} 
 & \textit{InterClass1:} BIRD (37,715) \\ \cline{2-2} 
 & \textit{InterClass2:} SIIM (33,125) \\ \cline{2-2} 
 & \textit{InterClass3:} IVC-Filter (1,258) \\ \hline
\multirow{5}{*}{IVC-Filter} & \textit{In-class:} type 11 (196) \\ \cline{2-2} 
 & \textit{Intra-class:} type 0-10, 12,13 (1,062) \\ \cline{2-2} 
 & \textit{InterClass1:} BIRD (37,715) \\ \cline{2-2} 
 & \textit{InterClass2:} SIIM (33,125) \\ \cline{2-2} 
 & \textit{InterClass3:} RSNA (30,227) \\ \hline
\multirow{5}{*}{SIIM} & \textit{In-class:} benign (32,541) \\ \cline{2-2} 
 & \textit{Intra-class:} malignant (584) \\ \cline{2-2} 
 & \textit{InterClass1:} BIRD (37,715) \\ \cline{2-2} 
 & \textit{InterClass2:} IVC-Filter (1,258) \\ \cline{2-2} 
 & \textit{InterClass3:} RSNA (30,227) \\ \hline
\end{tabular}%
}
\vspace{-3mm}
\end{table}

\begin{table*}[tp]
\caption{Intra-class OOD detection results (FPR, TPR and AUC values) of various anomaly detectors trained on RSNA, IVC-Filter and SIIM datasets. Best results are highlighted. Standard deviations are calculated via 10 rounds of bootstrapping estimations. }
\label{tab:cvae_intraclass}
\centering
\resizebox{\textwidth}{!}{%
\begin{tabular}{|l|ccc|ccc|ccc|}
\hline
\multirow{2}{*}{Methods}  & \multicolumn{3}{c|}{RSNA}  & \multicolumn{3}{c|}{IVC-Filter} &\multicolumn{3}{c|}{SIIM}  \\ \cline{2-10} 
  & $\downarrow$\textit{FPR}      & $\uparrow$\textit{TPR}    & $\uparrow$\textit{AUC}    
  & $\downarrow$\textit{FPR}      & $\uparrow$\textit{TPR}    & $\uparrow$\textit{AUC}   
  & $\downarrow$\textit{FPR}      & $\uparrow$\textit{TPR}    & $\uparrow$\textit{AUC} \\
  \hline \hline 
AE~\cite{sakurada2014anomaly}  & $0.318^{\pm0.014}$   & $0.461^{\pm0.009}$ & $0.566^{\pm0.004}$  & $0.198^{\pm0.104}$     & $0.350^{\pm0.075}$    & $0.436^{\pm0.040}$      & $0.420^{\pm0.024}$   & $0.714^{\pm0.030}$  & $0.673^{\pm0.006}$  \\
VAE~\cite{an2015variational}   & $0.473^{\pm0.001}$   & $0.462^{\pm0.001}$ & $0.487^{\pm0.001}$  & $0.489^{\pm0.097}$     & $0.707^{\pm0.076}$    & $0.542^{\pm0.080}$     & $0.442^{\pm0.008}$   & $0.740^{\pm0.006}$   & $0.676^{\pm0.023}$  \\
pchVAE~\cite{zimmerer2019high}   & $0.501^{\pm0.018}$   & $0.731^{\pm0.030}$ & $0.600^{\pm0.007}$   & $0.590^{\pm0.072}$     & $0.620^{\pm0.013}$   & $0.472^{\pm0.038}$      & $0.378^{\pm0.045}$   & $0.558^{\pm0.040}$  & $0.616^{\pm0.012}$\\
DeepSVDD~\cite{ruff2018deep}   & $0.508^{\pm0.021}$ & $0.413^{\pm0.023}$  & $0.421^{\pm0.009}$   & $0.503^{\pm0.106}$     & $0.672^{\pm0.042}$  & $0.500^{\pm0.075}$     & $0.276^{\pm0.036}$      & $0.683^{\pm0.050}$     & $0.740^{\pm0.010}$    \\
GANomaly~\cite{akcay2018ganomaly}   & $0.524^{\pm0.005}$   & $0.678^{\pm0.015}$  & $0.576^{\pm0.005}$    & $0.446^{\pm0.172}$     & $0.627^{\pm0.227}$    & $0.518^{\pm0.103}$   & $0.553^{\pm0.103}$      & $0.495^{\pm0.108}$     & $0.418^{\pm0.016}$  \\
f-AnoGAN~\cite{schlegl2019f}   & $0.365^{\pm0.033}$   & $0.541^{\pm0.029}$  & $0.614^{\pm0.005}$    & $0.419^{\pm0.077}$     & $0.611^{\pm0.054}$    & $0.544^{\pm0.042}$   & $0.381^{\pm0.000}$    & $0.624^{\pm0.033}$    & $0.721^{\pm0.015}$  \\
\hline
\textit{CVAD (ours)}    &\textit{0.327$^{\pm0.016}$}  & \textit{0.646$^{\pm0.017}$} & \textbf{\textit{0.696$^{\pm0.005}$}}  & \textit{0.541$^{\pm0.094}$}   & \textit{0.706$^{\pm0.091}$}  & \textbf{\textit{0.582$^{\pm0.031}$}}  & \textit{0.376$^{\pm0.020}$}   & \textit{0.766$^{\pm0.021}$}   & \textbf{\textit{0.749$^{\pm0.010}$}}\\
\hline
\end{tabular}%
}
\vspace{-3mm}
\end{table*}

\section{Experiments}
We conducted extensive experiments, verifying the generalizability and effectiveness of our approach on multiple open-access medical image datasets for intra- and inter-class OOD detection. In total, we used four independent datasets, including three medical image datasets -- RSNA Pneumonia dataset~\cite{wang2017chestx}, inferior vena cava filters (IVC-Filter in short) on radiographs~\cite{ni2020deep} and SIIM-ISIC Melanoma dataset~\cite{rotemberg2021patient} (identify melanoma in lesion images) and one natural image datasets -- Bird Species\footnote{\url{https://www.kaggle.com/gpiosenka/100-bird-species}}. Among the medical datasets, RSNA and SIIM datasets have binary classes -- normal and abnormal, whereas IVC-Filter dataset has 14 distinct types (classes). Table~\ref{details} lists the class information and number of images for each dataset and the corresponding usage in the \textbf{Detail} column. Bird dataset, which contains 270 bird species with 38,518 training images, was only used as inter-class OOD for detection validation. To unify the OOD detection pipeline and facilitate evaluation, we resized both the medical images and the validation inter-class OOD images to a unified $256\times256\times channel$ size, where IVC-Filter and RSNA datasets are in gray scale with $channel$ as 1 and the SIIM images are in RGB format and have $channel$ 3. To train the anomaly detectors, we split the ID data into training and valuation parts in the ratio of 80\% v.s. 20\%. All the OOD data will only be used during evaluation phase.

We implemented our model using Pytorch 1.5.0, Python 3.6. $\alpha_1, \alpha_2$ were equal to 1. We ran the models on 4 NVIDIA Quadro RTX 6000 GPUs with 24 GB memory each. In our model training, we used Adam optimizer with a learning rate of 0.001, and each network was trained for 100-350 epochs.

We evaluated our anomaly detection model performance in terms of standard statistical metrics - (i) area under the receiver operating characteristic (AUROC, AUC in short); (ii) True Positive rate (TPR); (iii) False positive rate (FPR). To classify ID and OOD classes, a threshold should be defined for the anomaly scores. Notably, the AUC value is threshold-invariant, while the TPR and FPR are determined by the selection of the anomaly threshold. We adopted the Geometric Mean (G-Mean) method to determine an optimal threshold for the ROC curve by tuning the decision thresholds and reported the resulting FPR and TPR values. To be fair and thorough, we ran all the experiments on both intra-class OOD and inter-class OOD to further analyze the performance of anomaly detectors on the specific type of OOD detection. 

\section{Results}
\subsection{Quantitative Results}
We set the vanilla AE and VAE architectures as baselines and compared our CVAD model with several representative models with varying architectures -- pchVAE~\cite{zimmerer2019high}, a classifier-based approach DeepSVDD~\cite{ruff2018deep}, and two GAN-based methods, i.e., GANomaly~\cite{akcay2018ganomaly} and f-AnoGAN~\cite{schlegl2019f}. Table~\ref{tab:cvae_intraclass} shows the models' performance for the intra-class OOD detection and Table~\ref{tab:cvae_interclass} primarily presents the inter-class OOD performance. 
\begin{table}[htp]
\caption{AUC scores predicted by OOD detectors for inter-class identification on RSNA, IVC-Filter and SIIM datasets. Bold indicates the best performance.}
\label{tab:cvae_interclass}
\resizebox{0.5\textwidth}{!}{
\begin{tabular}{|l|l|c|c|c|}
\hline
\multirow{2}{*}{\textbf{Dataset}}  & \multirow{2}{*}{\textbf{Methods}} & \multicolumn{3}{c|}{\textbf{AUROC score}} \\  
\cline{3-5} 
 &   & \textit{InterClass1} & \textit{InterClass2} & \textit{InterClass3} \\ \hline \hline
 \multirow{6}{*}{RSNA}  & AE~\cite{sakurada2014anomaly} & $0.677^{\pm0.006}$ & $0.608^{\pm0.005}$ & $0.616^{\pm0.004}$ \\ 
  & VAE~\cite{an2015variational} & $0.752^{\pm0.004}$ & $0.604^{\pm0.007}$ & $0.613^{\pm0.006}$ \\ 
  & pchVAE~\cite{zimmerer2019high} & $0.790^{\pm0.006}$  & $0.776^{\pm0.005}$ & $0.632^{\pm0.007}$ \\ 
  & DeepSVDD~\cite{ruff2018deep} & $0.838^{\pm0.005}$ & \textbf{0.834$^{\pm0.004}$} & $0.604^{\pm0.006}$ \\
  & GANomaly~\cite{akcay2018ganomaly} & $0.733^{\pm0.005}$ & $0.816^{\pm0.004}$ & $0.597^{\pm0.007}$ \\ 
  & f-AnoGAN~\cite{schlegl2019f} & $0.842^{\pm0.001}$ & $0.693^{\pm0.001}$ & $0.682^{\pm0.002}$ \\ 
  & \textit{CVAD (ours)} & \textbf{\textit{0.863$^{\pm0.003}$}} & \textit{0.803$^{\pm0.004}$} & \textbf{\textit{0.703$^{\pm0.005}$}} \\ \hline\hline

\multirow{6}{*}{IVC-Filter} & AE~\cite{sakurada2014anomaly} & $0.372^{\pm0.051}$ & $0.342^{\pm0.041}$ & $0.237^{\pm0.051}$ \\ 
  & VAE~\cite{an2015variational} & $0.666^{\pm0.026}$ & $0.400^{\pm0.039}$ & $0.706^{\pm0.027}$ \\ 
  & pchVAE~\cite{zimmerer2019high} & $0.885^{\pm0.022}$ & $0.732^{\pm0.033}$ & $0.905^{\pm0.026}$ \\ 
  & DeepSVDD~\cite{ruff2018deep} & $0.861^{\pm0.051}$ & $0.724^{\pm0.060}$ & $0.883^{\pm0.102}$ \\ 
  & GANomaly~\cite{akcay2018ganomaly} & $0.803^{\pm0.018}$ & $0.827^{\pm0.190}$ & $0.922^{\pm0.072}$ \\ 
  & f-AnoGAN~\cite{schlegl2019f} & $0.911^{\pm0.020}$ & $0.625^{\pm0.043}$ & $0.864^{\pm0.042}$ \\ 
  & \textit{CVAD (ours)} & \textbf{\textit{0.984$^{\pm0.002}$}} & \textit{\textbf{0.911}$^{\pm0.017}$} & \textit{\textbf{0.985}$^{\pm0.001}$} \\ \hline \hline
 
\multirow{6}{*}{SIIM}& AE & $0.572^{\pm0.004}$ & $0.013^{\pm0.000}$ & $0.752^{\pm0.005}$ \\ 
  & VAE~\cite{an2015variational} & $0.712^{\pm0.006}$ & $0.021^{\pm0.002}$ & $0.759^{\pm0.003}$ \\   
  & pchVAE~\cite{zimmerer2019high} & $0.943^{\pm0.002}$ & \textbf{0.992}$^{\pm0.000}$ & $0.684^{\pm0.004}$ \\ 
  & DeepSVDD~\cite{ruff2018deep} & $0.980^{\pm0.001}$ & \textbf{0.992}$^{\pm0.000}$ & $0.804^{\pm0.002}$ \\ 
  & GANomaly~\cite{akcay2018ganomaly} & $0.688^{\pm0.005}$ & $0.989^{\pm0.000}$ & $0.442^{\pm0.006}$ \\ 
  & f-AnoGAN~\cite{schlegl2019f} & $0.951^{\pm0.001}$ & $0.924^{\pm0.002}$ & $0.606^{\pm0.003}$ \\ 
  & \textit{CVAD (ours)} & \textbf{\textit{0.983$^{\pm0.001}$}} & \textit{0.978$^{\pm0.001}$} & \textbf{\textit{0.869$^{\pm0.003}$}} \\ \hline 
\end{tabular}
}
\end{table}

\subsubsection{Results for Intra-class OOD Detection} 
Intra-class OOD images are the most challenging outliers to identify since they often share similarity to the ID data but belong to a different class with unique characteristics. 
Still, CVAD exhibits its superiority in detecting intra-class OOD for medical images. On the RSNA dataset, CVAD achieves the best AUC score 0.696 (+0.275 from DeepSVDD's AUC score 0.421, +0.120 from GANomaly's AUC score 0.576, +0.082 from f-AnoGAN's AUC score 0.614); for IVC-Filter, CVAD obtains the highest AUC values 0.582; for SIIM dataset, although DeepSVDD and f-AnoGAN show competitive performance, CVAD acquires the optimal AUC score 0.749. Overall, CVAD performs stably and effectively for intra-class OOD detection.

\subsubsection{Results for Inter-class OOD Detection}
To fairly evaluate all the models, we tested them on multiple inter-class OOD data types and presented the corresponding AUC scores in Table.~\ref{tab:cvae_interclass}. As the OOD image datasets may have different image channels and image sizes from the ID training images, we adjusted the image channels and resized the images to ensure consistent input data format for evaluation\footnote{\small{For example, to evaluate trained models on RSNA, we converted the BIRD and SIIM images to grayscale mode and resized them to the same in-distribution image size.}}. CVAD obtains the highest AUC values on RSNA and SIIM datasets (except for inter-class2), and performs the best for IVC-Filter dataset across three inter-class OOD detection evaluations. Generally, the inter-class OOD detection of CVAD is satisfied with stable performance.

\begin{table}[tp]
\caption{AUC scores predicted by the ``generator" CVAD\_G, the discriminator CVAD\_D and CVAD for inter-class identification on RSNA, IVC-Filter and SIIM datasets respectively. }
\label{tab:cvae_interclass_ab}
\resizebox{0.5\textwidth}{!}{
\begin{tabular}{|l|l|c|c|c|c|}
\hline
\multirow{2}{*}{\textbf{Dataset}}  & \multirow{2}{*}{\textbf{Methods}} & \multicolumn{4}{c|}{\textbf{AUROC score}} \\  
\cline{3-6} 
 &   & \textit{IntraClass} & \textit{InterClass1} & \textit{InterClass2} & \textit{InterClass3} \\ \hline \hline
 \multirow{3}{*}{RSNA} 
& \textit{CVAD\_G (ours)} & $0.602^{\pm0.006}$   & $0.854^{\pm0.003}$ & $0.517^{\pm0.004}$ & $0.601^{\pm0.005}$ \\ 
& \textit{CVAD\_D (ours)} & $0.672^{\pm0.005}$  & $0.793^{\pm0.003}$ & $0.809^{\pm0.003}$ & $0.679^{\pm0.005}$ \\ 
& \textit{CVAD (ours)} & 0.696$^{\pm0.005}$ & 0.863$^{\pm0.003}$ & 0.803$^{\pm0.004}$ & 0.703$^{\pm0.005}$ \\ \hline\hline

\multirow{3}{*}{IVC-Filter} 
& \textit{CVAD\_G (ours)} & $0.568^{\pm0.031}$  & 0.981$^{\pm0.003}$ & 0.787$^{\pm0.023}$ & 0.983$^{\pm0.002}$ \\
& \textit{CVAD\_D (ours)} & $0.543^{\pm0.041}$ & 0.661$^{\pm0.018}$ & 0.925$^{\pm0.011}$ & 0.834$^{\pm0.013}$ \\
& \textit{CVAD (ours)} & $0.582^{\pm0.031}$ & 0.984$^{\pm0.002}$ & 0.911$^{\pm0.017}$ & 0.985$^{\pm0.001}$ \\ \hline \hline
 
\multirow{3}{*}{SIIM}
& \textit{CVAD\_G (ours)} &  $0.746^{\pm0.010}$ & $0.995^{\pm0.000}$ & $0.995^{\pm0.000}$ & $0.827^{\pm0.004}$ \\ 
& \textit{CVAD\_D (ours)} & $0.724^{\pm0.008}$ & $0.874^{\pm0.002}$ & $0.055^{\pm0.001}$ & $0.862^{\pm0.005}$ \\ 
& \textit{CVAD (ours)} & $0.749^{\pm0.010}$ & 0.983$^{\pm0.001}$ & 0.978$^{\pm0.001}$ & 0.869$^{\pm0.003}$ \\ \hline 
\end{tabular}
}
\end{table}

\subsubsection{Effectiveness of CVAD's Components}
We here demonstrate the importance of each component of CVAD. Table~\ref{tab:cvae_interclass_ab} shows the performance difference under the intra-class and three inter-class OOD data situations. CVAD\_G represents the ``generator", CVAD\_D stands for only using the predictions of the discriminator. CVAD balances the two components' prediction. As can be observed, CVAD\_G and CVAD\_D show certain variations for different cases. For example, CVAD\_D generally works better than CVAD\_G for RSNA dataset but behaves worse than CVAD\_G in SIIM scenario. Nevertheless, each component owns its unique OOD discriminative ability, and combining their advantages entitles CVAD the capability of capturing both intra-class and inter-class dissimilarities. For which sake, CVAD has better generalization and can perform well and stably under different situations. 


\subsection{Qualitative Results}
\subsubsection{Anomaly Detection}
\begin{figure}[tp]
\centering
\includegraphics[width=0.48\textwidth]{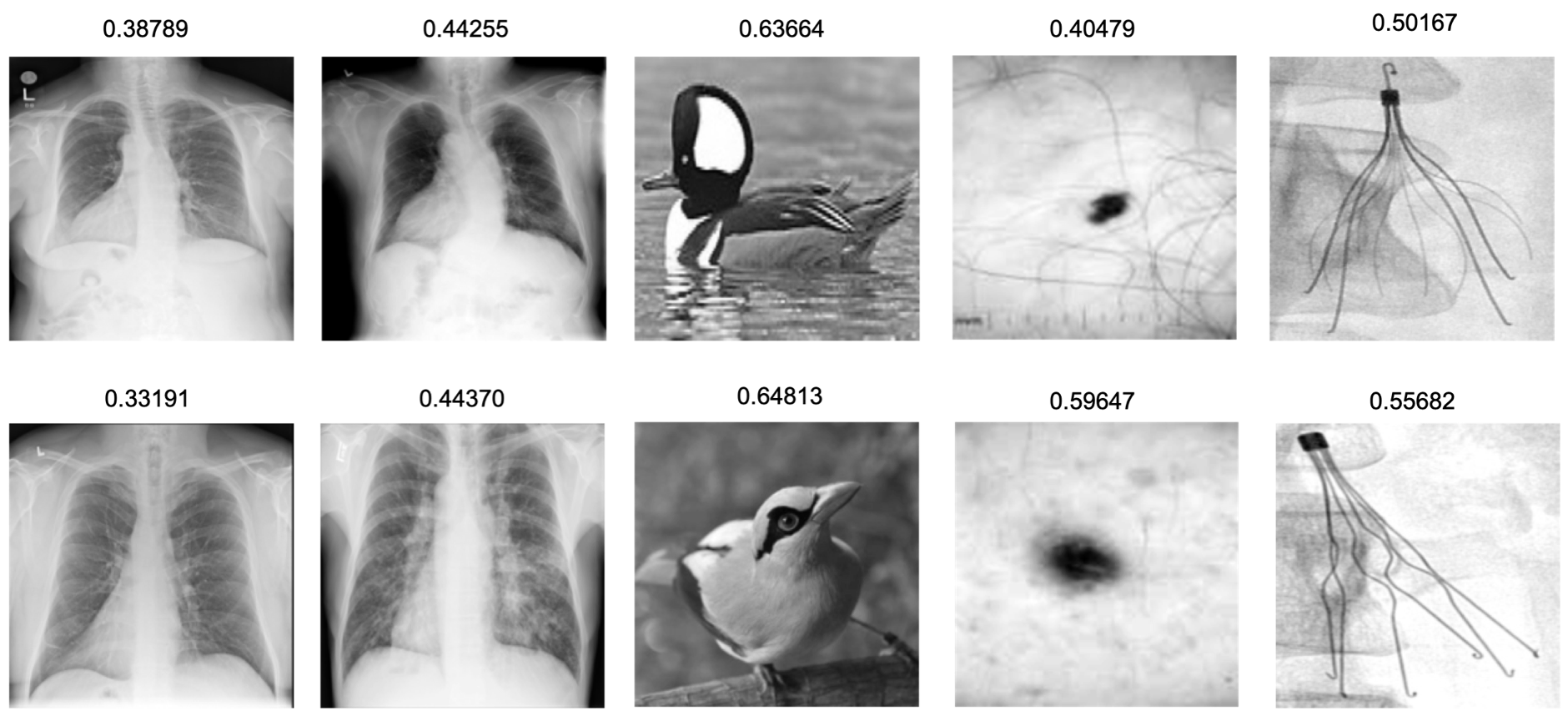}
\caption{Anomaly scores output by CVAD for different types of input data (experiments for RNSA dataset). Columns from left to right, ID, intra-class OOD, inter-class OOD1, inter-class OOD2, inter-class OOD3.}
\vspace{-2mm}
\label{rsna_examples}
\end{figure}

Figure~\ref{rsna_examples} shows two experimental results for RSNA dataset. Each row stands for one case and each column represents a specific type of input data. From left to right, they are in-distribution data, intra-class OOD data, inter-class OOD1 data, inter-class OOD2 data and inter-class OOD3 data, respectively. The corresponding anomaly score predicted by CVAD is on top of each example. Higher anomaly scores mean more likely the inputs are OOD. As can be seen in Figure~\ref{rsna_examples}, the two intra-class OOD samples (2nd column) are alike as the in-distribution data but the inter-class OOD examples show very different appearance from in-distribution data. Correspondingly, the anomaly scores of intra-class OOD are close to the scores of ID samples and difficult to separate whereas the intra-class OOD cases with clear variations are assigned higher anomaly scores and are easy to identify. This phenomenon further demonstrates the challenges of identifying intra-class OOD data.

\begin{figure}[t]
\includegraphics[width=0.45\textwidth]{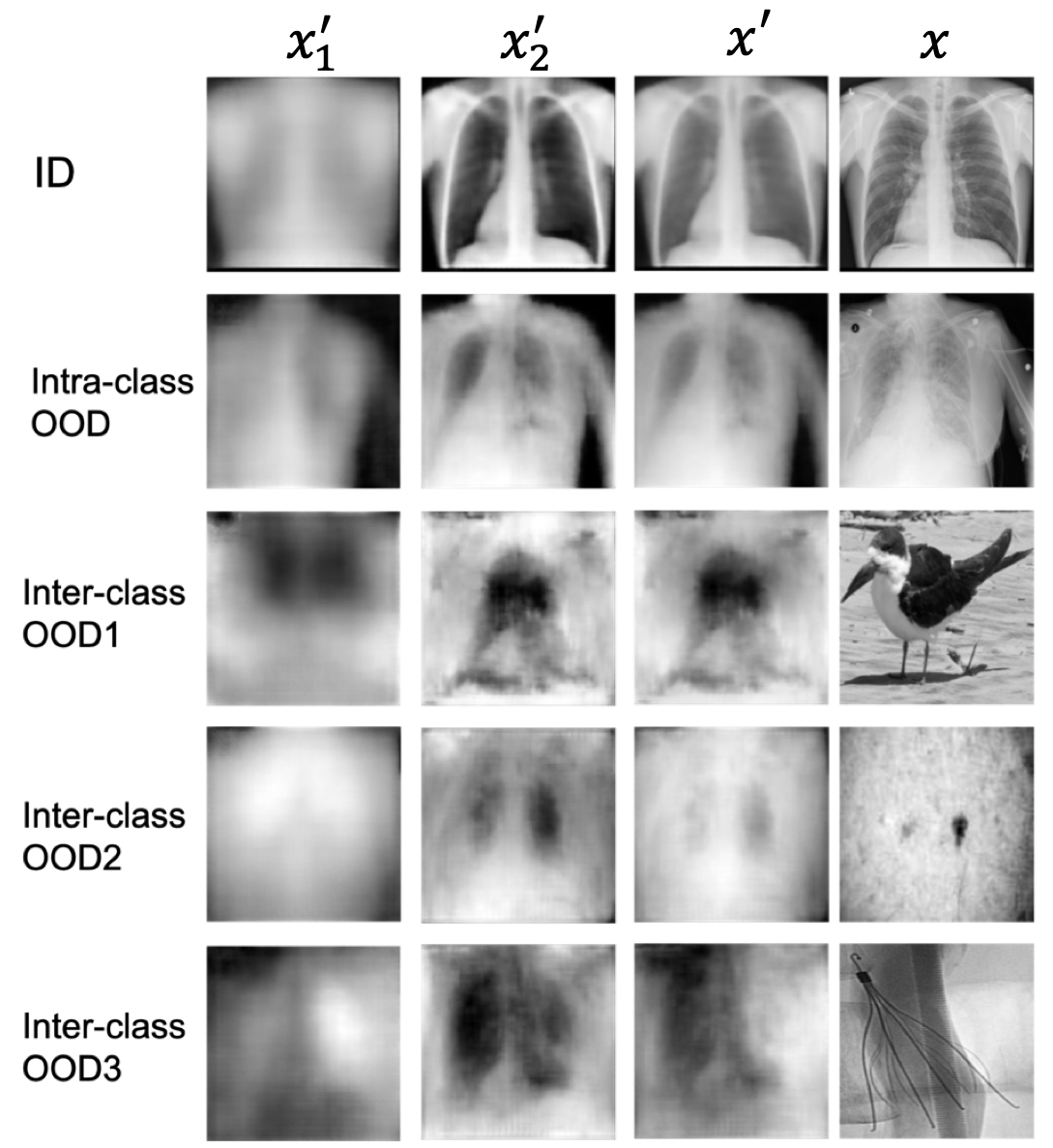}
\caption{Reconstruction details visualization of CVAD's ``generator" trained on RSNA dataset for different data types.}
\label{rsna}
\vspace{-5mm}
\end{figure}

\subsubsection{Visualization of Reconstruction Effects}
CVAD gains good latent in-distribution features via its ``generator", which learns both low-level and high-level representations. To demonstrate the effectiveness, we took RSNA dataset as a representative and showcased the reconstruction details in Figure~\ref{rsna}, with the first column for branch VAE reconstruction $x_{2}'$, the second column for standard VAE part reconstruction $x_{1}'$, the third column for ultimate reconstruction $x'$ and the last column for the original input image $x$ (following the same notations indicated in Figure~\ref{fig_cvad}). To further reveal the effects of ``generator" on different OOD samples, we also presented example images for ID (i.e., normal class, 1st row), intra-class OOD (i.e., pneumonia or with opacity, 2nd row), inter-class OOD1 (i.e., gray-scale bird images, 3rd row), inter-class OOD2 (i.e., skin cancer images from SIIM dataset,4th row) and inter-class OOD3 (i.e., images from IVC-Filter dataset, 5th row) in Figure~\ref{rsna}. Compared with the intra-class medical OOD data, reconstructions on inter-class OOD inputs are more messy and dissimilar to the original OOD data, which leads to larger reconstruction errors and thus easier to distinguish. This observation reveals the varying difficulties of detecting different types of OOD data -- intra-class OOD is much more challenging than inter-class OOD.

\section{Conclusion}
We propose an effective medical anomaly detector CVAD that can reconstruct coarse and fine image components by learning multi-scale latent representations. The high quality of generated images enhances the discriminative ability of the binary discriminator in identifying unknown OOD data. We demonstrate the OOD detection efficacy for both intra-class and inter-class OOD data on various medical and natural image datasets. Our model has no prior assumptions on the input images and application scenarios for OOD, thus can be applied to detect OOD samples in a generic way for multiple scenarios. A detailed technical report about the code implementation and parameter usages of CVAD has been publicly available for easy reproduction.

\clearpage






%

\end{document}